# Star-forming gas in young clusters

**Philip C. Myers**

Harvard-Smithsonian Center for Astrophysics, 60 Garden Street, Cambridge MA 02138 USA

pmyers@cfa.harvard.edu

**Abstract.** Initial conditions for star formation in clusters are estimated for protostars whose masses follow the initial mass function (IMF) from 0.05 to 10 solar masses. Star-forming infall is assumed equally likely to stop at any moment, due to gas dispersal dominated by stellar feedback. For spherical infall, the typical initial condensation must have a steep density gradient, as in low-mass cores, surrounded by a shallower gradient, as in the clumps around cores. These properties match observed column densities in cluster-forming regions when the mean infall stopping time is 0.05 Myr and the accretion efficiency is 0.5. The infall duration increases with final protostar mass, from 0.01 to 0.3 Myr, and the mass accretion rate increases from 3 to 300 × $10^{-6}$ solar masses/yr. The typical spherical accretion luminosity is ~5 solar luminosities, reducing the "luminosity problem" to a factor ~3. The initial condensation density gradient changes from steep to shallow at radius 0.04 pc, enclosing 0.9 solar masses, with mean column density 2 × $10^{22}$ cm$^{-2}$, and with effective central temperature 16 K. These initial conditions are denser and warmer than those for isolated star formation.

*Key words:* ISM: clouds - stars: formation

# 1. Introduction

Most stars form in clusters (Lada & Lada 2003), yet the initial conditions for such star formation are poorly known, and it is not clear how they differ from the better known initial conditions for isolated star formation.

Many young clusters have been studied recently, at mm, sub-mm, and infrared wavelengths. They are generally found in parsec-scale regions of high column density, which in turn have prominent filamentary substructure and numerous embedded cores (Walsh et al 2007, Nutter, Ward-Thompson & André 2006, Wilking, Gagné, & Allen 2008, Gutermuth et al 2009, Myers 2009a).

However, observations of young clusters do not yield estimates of initial properties as easily as do observations of isolated cores. Cores in young clusters are much closer to each other, and to already formed young stars, than are starless cores in isolated regions (Lada, Strom, & Myers 1993). Many cores in young clusters appear blended, and knowledge of their structure is limited by confusion and insufficient resolution. Young clusters and their cores have been modified by outflows from nearby protostars (Bally et al 1999, Sandell & Knee 2001, Stanke & Williams 2007). As a result the initial structure of the star-forming gas in clusters is still unclear.

This paper explores a new way to infer the properties of star-forming gas in clusters. The basic idea is that the final masses of protostars in a cluster follow the initial mass function of stars (IMF). Then the IMF and an assumed distribution of infall durations together specify the initial density structure. Relations among the distribution of infall durations, the initial density structure, and the protostar mass function were presented in Myers (2009b, hereafter Paper 1).

In Paper 1, the initial structure was assumed to be one of several "core-environment" systems. Stellar feedback and other asynchronous ways to disperse dense gas were assumed to terminate infall with equal likelihood at any moment. The infall duration followed an exponential waiting-time distribution. Each resulting protostar mass distribution resembled the initial mass function of stars (IMF), although no best-fit model was identified.

This paper takes the IMF as a given property, and derives the unique initial density profile of a condensation which can form a star of any mass, depending on its infall duration. The same distribution of infall durations is assumed as in Paper 1. To account for uncertainty in the IMF, two versions are used (Kroupa 2002, Chabrier 2005), but these give essentially the same result.

The inferred initial structure has a steep density gradient on small scales, resembling that of isolated dense cores described by isothermal models. This structure makes a smooth transition to lower-density gas having a shallower gradient, as is observed in more turbulent "clumps." The same model also predicts the protostar formation time, mass accretion rate, and accretion luminosity as functions of protostar mass.

The paper is organized into five sections. Section 2 gives the basic assumptions and equations of the model. Section 3 presents results on infall duration, mass accretion rate, and density structure. Section 4 discusses the results, and Section 5 gives a summary of the paper.

## 2. Model

The basic premise is that the final protostar mass $M_\star$ increases monotonically with increasing infall duration $t_f$, and that $M_\star$ does not depend significantly on any independent variable other than $t_f$. Consequently the probability densities for $M_\star$ and for $t_f$ have a well-defined relationship. Section 2.1 gives the underlying assumptions, and Section 2.2 relates probability densities for $M_\star$ and $t_f$. Sections 2.3-2.5 derive the resulting expressions for infall duration, mass accretion rate, accretion luminosity and density of the initial gas. Section 2.6 discusses the values adopted for the two adjustable parameters.

*2.1. Definitions and assumptions*

*2.1.1. Constant accretion efficiency* In this model, a "protostar" is a single young stellar object (YSO) which has not yet reached the main sequence. The mass of its accretion disk is assumed to be negligible, since the ratio of disk mass to YSO mass is typically 0.1 in nearby star-forming regions (Andrews & Williams 2007, Jørgensen et al 2009). Binaries and higher multiples are ignored. This single protostar forms at the center of a spherically symmetric "condensation." This general term is used because the condensation has properties of both a "core" and a "clump."

It is assumed that the final protostar mass $M_\star$ forms in time $t_f$, and that $M_\star$ scales with the initial gas mass $M$ available for infall in a free-fall time $t_f$, according to

$$M_\star = \varepsilon M \quad . \qquad (1)$$

Here the mass accretion efficiency $\varepsilon$ is a parameter independent of time and is constant from core to core. The departure of this accretion efficiency from unity reflects the difference between the protostar mass accreted in time $t_f$ and that predicted by the idealized model of cold steady spherical infall from rest. In more detailed models this difference is usually a factor of order 2 (Shu 1977, Terebey, Shu & Cassen 1984, Fatuzzo, Adams & Myers 2004). The approximation that $\varepsilon$ is constant may be least accurate for small masses and early times (Paper 1).

If $\varepsilon$ is not constant but follows a distribution, the resulting distribution of $M_\star$ is the convolution of the distributions of $\varepsilon$ and $M$, and is broader than the distribution of $M$ alone. This difference in width is relatively small provided the distribution of $\varepsilon$ is much narrower than the distribution of $M$. As an example, the greater and lesser half-maximum (HM) masses of the IMF of Kroupa (2002) have ratio 18. The broadening of such a distribution due to convolution with another distribution having HM mass ratio of 3 can be judged from Figure 8 of Paper 1. There the convolved distribution of log mass is broader than the unconvolved distribution by a factor 1.12. For this paper, such an increase due to distributed values of $\varepsilon$ is considered negligible.

The protostars considered here have completed their accretion, and their masses are "final" masses. An ensemble of such protostars approximates a young cluster which has already formed most of its members, and for which most members are no longer gaining significant mass. Such an ensemble is common among 36 "embedded" clusters within 1 kpc, where more than 80% of the members typically belong to evolutionary class II or III (Gutermuth et al 2009). The mass distributions considered here therefore do not refer to the youngest clusters, whose mass functions are still evolving. Such systems have been discussed by Fletcher & Stahler (1994a, b) and by McKee & Offner (2009).

*2.1.2. Centrally condensed gas*  The initial condensation is assumed to be sufficiently concentrated so that its mass $M$ within a given radius increases monotonically with the free fall time $t_f$ of gas within the same radius. This assumption is justified because $M$ increases monotonically with increasing $t_f$ for many different density profiles. From the definition of the free-fall time (Hunter 1962), $M$ increases monotonically with $t_f$ if the mean density within a given radius decreases outward, and if the enclosed mass increases outward. If the mean density varies as radius to the power $-p$, then $M$ increases monotonically with $t_f$ for $0 < p < 3$, encompassing many descriptions of core and clump structure (Stüwe 1990, Bergin & Tafalla 2007). For such centrally condensed structures, equation (1) indicates that $M_\star$ also increases monotonically with $t_f$.

2.1.3. *Similar initial condensations*  In a young cluster, condensations which make one protostar each are assumed to have sufficiently similar dependence of mass on radius so that a "typical" density structure is meaningful. This assumption is justified by the narrow range of temperatures of star-forming dense cores. For example the kinetic temperature of 36 dense cores in the Orion A region is distributed with HM values 15 K and 24 K, according to observations of the (1,1) and (2,2) lines of $NH_3$ (Jijina, Myers & Adams 1999). A similar result was reported by Li, Goldsmith & Menten (2003). The corresponding distribution of Jeans masses has HM values with ratio of ~2, assuming constant peak density. Allowing for variation in peak density by a factor of a few increases the likely ratio of HM values to ~3. This ratio is small compared to the ratio ~18 of HM values of the IMF. Therefore, as discussed above for distributed accretion efficiency, these condensation masses are similar enough to allow the assumption that they are identical, with relatively little error.

Such a distribution of similar star-forming condensation masses is not necessarily inconsistent with the broad distribution of observed core masses reported by many authors, including Motte, André & Neri (1998), and Alves, Lombardi & Lada (2007). As discussed in Paper 1, not all observed starless cores make single stars. Some observed cores have too little mass and are too lightly bound to form any protostars before they are dispersed (Enoch et al 2008). Other observed cores are more massive and are likely to form many stars, as in B59 (Brooke et al 2007). Similar cautions against a one-to-one interpretation of observed core masses and protostar masses are given by Goodwin et al (2008), Hatchell & Fuller (2008), and Swift & Williams (2008).

The foregoing evidence for similarity of the initial mass profile over the ensemble of condensations implies that the typical protostar mass does not depend significantly on independent variables other than the infall duration.

*2.2. Probability densities*

Following the above assumptions, the probability that the final protostar mass lies between $M_\star$ and $M_\star + dM_\star$ is equal to the probability that its infall duration lies between $t_f$ and $t_f + dt_f$, or

$$p(M_\star)dM_\star = p(t_f)dt_f . \qquad (2)$$

This relationship of probability densities is well-defined, due to the monotonic dependence of $M_\star$ on $t_f$, and to the dependence of $M_\star$ only on $t_f$. If the protostar mass depends significantly on additional initial properties, then equation (2) must be replaced by a more detailed treatment, as in Section 3.4 of Paper 1.

*2.2.1. Protostar masses* The probability density $p(M_\star)$ is related to the mass function, denoted $\Phi$, by

$$\Phi \equiv M_\star p(M_\star) . \qquad (3)$$

It is assumed that the final masses of protostars in young clusters have mass function $\Phi$ and probability density $p(M_*)$ set by the IMF. Two descriptions of the IMF are used here, to represent the IMF and its uncertainty. Each has a local maximum at the modal mass, near 0.2 $M_\odot$, and a decreasing power-law segment at high mass, matching that of Salpeter (1955). At lower mass, the IMF of Kroupa (2002) has two power-law segments and the IMF of Chabrier (2005) has a log-normal segment. These segmented functions are approximated by smooth analytic functions having continuous derivatives, given in the Appendix.

The probability densities in equations (A1) and (A3) are combined with equation (3) to give approximations to the IMF, denoted $\Phi_K$ and $\Phi_C$. These fit their corresponding segmented functions well enough to follow their general shapes and to represent their differences, as shown in Figure 1. In relation to $\Phi_K$, $\Phi_C$ peaks at slightly higher mass and has a steeper low-mass slope.

2.2.2. *Infall durations*. It is assumed that the final protostar mass is associated with a well-defined infall duration. This assumption assigns a single time $t_f$ to the more complex history of the accretion, which is likely intermittent (Vorobyov & Basu 2005) and gradually decreasing (Bontemps et al 1996). The probability density of the infall duration in equation (2) can be written as the probability $E$ that the infall endures until $t_f$, times the probability density $s$ that the infall stops at $t_f$,

$$p(t_f) = E(t_f)s(t_f) \quad . \qquad (4)$$

It is assumed that the infall is equally likely to stop at any moment, so that $s(t_f)$ is independent of $t_f$. As explained below, this assumption is justified in cluster-forming regions where gas dispersal is dominated by stellar feedback and other asynchronous causes.

Outflow from a protostar is the best-known way to disperse nearby dense gas, over the first few 0.1 Myr of protostar age (Arce et al 2007). But observed outflows appear too collimated to completely terminate the infall onto their protostars. The lowest-mass stars and brown dwarfs require infall duration < 0.1 Myr for standard infall models (Reipurth & Clarke 2001). Observed outflows from protostars with such young age estimates are jetlike, and remove only a small fraction of their dense core gas. Outflow opening angles increase with protostar age (Arce & Sargent 2006), but even the widest observed opening angles still leave a substantial mass of dense gas in equatorial latitudes (Qiu et al 2009). Despite its removal of dense gas, outflow from a protostar seems insufficient to provide the sole way to stop its infall.

In a young cluster, several additional ways are available to disperse dense gas. The stellar feedback due to outflows, heating, and ionization from nearby protostars may be more effective in dispersing the gas around a protostar than is its own solitary outflow. An example of multiple outflows disrupting a core with multiple protostars is in L1641N (Stanke & Williams 2007). Similarly, competitive accretion against nearby stars and cores, turbulent flows, and dynamical ejection can also remove dense gas from the neighborhood of a protostar. These external agents of dispersal are asynchronous: they operate at times which are independent of the start of the protostellar infall (Paper 1).

It remains to define quantitatively the physical conditions within young clusters, where such asynchronous dispersal becomes the dominant limitation on star-forming infall. The dispersal of a star-forming condensation by a neighbor outflow is most effective when substantial condensation gas lies within an outflow cone of a neighbor protostar, when the outflow momentum is sufficiently great, and when the condensation-neighbor separation is sufficiently small. In turn, the outflow momentum and necessary separation depend on both the mass and the age of the protostar. Detailed studies of such dispersal processes are needed, but are beyond the scope of this paper. Nonetheless it seems clear that dispersal is more effective, and is more asynchronous, as the density of neighbor protostars increases.

If such dispersal is most effective in crowded parts of clusters, it is possible that this process will apply only to a minority of the cluster stars, especially if the density of young stars falls off from a single central maximum. However, imaging studies suggest that in their early stages, nearby clusters have complex filamentary structure with several local maxima of protostar density (Gutermuth et al 2009). If so, the fraction of young cluster stars which are sufficiently crowded may be greater than expected from a space distribution with a single local maximum. It will be important to test this issue with realistic models of the space distribution of cluster protostars, in simulations of cluster evolution.

Assuming equally likely stopping, the probability density $s(t_f)$ that the infall stops at time $t_f$ in equation (4) is independent of how much time has already elapsed. The consequent distribution of infall durations is a waiting-time distribution whose form is the exponential function,

$$p(t_f) = \frac{1}{\bar{t}_f} \exp(-\theta) \qquad (5)$$

where $\bar{t}_f$ is the mean infall duration and where the dimensionless infall duration is

$$\theta \equiv \frac{t_f}{\bar{t}_f} \qquad (6)$$

(Feller 1968, Basu & Jones 2004, Nadarajah 2007, Paper 1). The exponential function is "memoryless" in that the probability of dispersal between times $\theta_1$ and $\theta_2$ is independent of how much time has elapsed before $\theta_1$. The time variation in equation (5) is therefore due only to the monotonically decreasing probability $E$ that the infall endures until $t_f$.

*2.3. Infall duration and mass accretion rate*

The probability $Q(M_\star)$ that the protostar mass exceed $M_\star$ decreases with increasing $M_\star$ and with increasing $\theta$. Integrating equation (2) and substituting $p(t_f)$ from equation (5) give

$$\theta = -\ln Q(M_\star) \qquad (7)$$

where

$$Q(M_\star) \equiv 1 - \int_0^{M_\star} dM'_\star \, p(M'_\star). \qquad (8)$$

Then equations (6)-(8) give the infall duration $t_f$ for a protostar of mass $M_\star$.

The instantaneous steady mass accretion rate at the time when the protostar has reached its final mass is obtained from equations (2), (6), (7), and (8) as

$$\frac{dM_\star}{dt} = \frac{Q(M_\star)}{\bar{t}_f \, p(M_\star)}. \qquad (9)$$

This mass accretion rate is a function only of protostar mass, and its only adjustable parameter is the mean infall time. If the accretion is nonsteady (Kenyon & Hartmann 1995, Vorobyov & Basu 2005, Baraffe, Chabrier, & Gallardo 2009), equation (9) represents a local average over a cycle of high and low accretion.

The mass accretion rate in equation (9) is nearly independent of protostar mass at low mass, and approaches a linear dependence at high mass. This property can be understood for protostar masses much less than, and much greater than the modal mass of the IMF. For low mass, the IMFs of Kroupa (2002) and Chabrier (2005) can be approximated by $\Phi \propto M_\star$, as is evident from Figure 1. Then equations (3), (8), and (9) give a constant mass accretion rate independent of protostar mass. For protostar mass significantly greater than the modal mass, the power-law nature of the IMF and equation (9) lead to a mass accretion rate proportional to the first power of protostar mass, $dM_\star/dt = M_\star/(s\,\bar{t}_f)$, where $s=1.35$ is the log-log slope of the high-mass IMF (Salpeter 1955). In terms of the IMF expression given in equation (A1), $s=b+c-1$.

*2.4. Accretion luminosity*

The accretion luminosity is one of numerous sources of radiative luminosity from a YSO (Adams & Shu 1986). It is assumed that the accretion luminosity is dominant among these sources for the Class 0 and Class I protostellar phases, as recently discussed by Dunham et al (2010).

The luminosity due to accretion onto the protostar can be written

$$L = \gamma L_s \qquad (10)$$

where $L_s$ is the luminosity due to accretion onto a spherical surface of radius $R_\star$

$$L_s = \frac{GM_\star dM_\star/dt}{R_\star}, \qquad (11)$$

and where the "accretion luminosity efficiency" $\gamma$ represents the departure of the accretion luminosity from that for purely spherical accretion. This efficiency is denoted $\varepsilon(1-\alpha)$ in equation (2) of Baraffe, Chabrier & Gallardo (2009, hereafter BCG). Here $\varepsilon$ is the ratio of internal to gravitational energy of the accreting material, which is $\leq 1/2$ for accretion from a thin disk at the protostar equator. The quantity $\alpha$ is the fraction of accreting internal energy absorbed by the protostar, where $\alpha \leq 1$. The radius $R_\star$ increases with mass and accretion rate, from 2 to 20 $R_\odot$ from the lowest to highest combinations of $M_\star$ and $dM_\star/dt$ considered here (Hosokawa & Omukai 2009). A typical value is 3 $R_\odot$ for $dM_\star/dt = 10^{-5}$ $M_\odot$ yr$^{-1}$ and for $M_\star = 0.1$-$1$ $M_\odot$, as also found by Stahler (1988). The accretion luminosity is discussed further in Sections 3.4 and 4.2.

## 2.5. Density and column density

The dimensionless infall duration $\theta$ is obtained from the definition of the free-fall time and from equations (1) and (6) as

$$\theta = \frac{\pi}{\bar{t}_f}\left(\frac{\varepsilon r^3}{8GM_\star}\right)^{1/2} \qquad (12)$$

where $r$ is the spherical radius enclosing the initial mass $M$, and where $G$ is the gravitational constant. Eliminating $\theta$ from equations (7) and (12) relates the initial radius and the final protostar mass,

$$\xi = \mu^{1/3}\left[-\ln Q(M_\star)\right]^{2/3} \qquad (13)$$

where the dimensionless mass $\mu$ is defined in equation (A2), where the dimensionless radius is defined by

$$\xi \equiv \frac{r}{r_0}, \qquad (14)$$

and where the radius scale is

$$r_0 \equiv 2\left(\frac{GM_{\star n}}{\varepsilon}\right)^{1/3}\left(\frac{\bar{t}_f}{\pi}\right)^{2/3}. \qquad (15)$$

The density is obtained from the derivative of mass with respect to radius (e.g. Shu 1992, equation 5.10), using equations (1) and (14), as

$$n = n_0\left(\xi^2 \frac{d\xi}{d\mu}\right)^{-1} \qquad (16)$$

where the density scale is

$$n_0 \equiv \frac{\pi}{32mG\bar{t}_f^2} \qquad (17)$$

where $m$ is the mean mass per gas particle. Evidently $n_0$ is one-third of the density whose free-fall time is $\bar{t}_f$.

The mean column density within a given radius is obtained from equations (1), (2), (14), and (15) as

$$\bar{N} = N_0 \frac{\mu}{\xi^2} \qquad (18)$$

with column density scale

$$N_0 \equiv \frac{(\pi M_{\star n}/\varepsilon)^{1/3}}{4mG^{2/3}\bar{t}_f^{4/3}} \qquad . \qquad (19)$$

## 2.6. Parameter values

The mass accretion model has two adustable parameters, the mass accretion efficiency $\varepsilon$ and the mean infall time $\bar{t}_f$. These parameters are constrained by the mean column density of the gas in young clusters. In a recent study, seven young clusters within 1 kpc have a "core" radius 0.2 pc enclosing the most crowded part of the cluster, with mean column density $1 \times 10^{22}$ cm$^{-2}$ (Gutermuth et al 2009). It is assumed that this core radius is comparable to the radius of the largest initial condensation. The mean column density $\bar{N}$ given by equations (18) and (19) matches the typical observed value at radius 0.2 pc with the choices $\varepsilon = 0.5$ and $\bar{t}_f = 0.05$ Myr. These parameter values are adopted for the calculations presented in the rest of this paper.

These parameter values are constrained to a relatively narrow range. The mean column density scales as $\varepsilon^{-1/3}(\bar{t}_f)^{-4/3}$, so keeping the mean column density constant while changing $\varepsilon$ by a factor 2 would require a corresponding change in $\bar{t}_f$ by a relatively small factor of $2^{-1/4}$. Regions with greater mean column density require reduction in $\varepsilon$, in $\bar{t}_f$, or in both. The adopted mean infall time corresponds to a typical mean density of star-forming gas $5 \times 10^5$ cm$^{-3}$.

## 3. Results

This section describes properties of the initial gas and the infall duration, calculated from the foregoing model, and compares them with available observational results.

*3.1. Protostar mass range*

The range of protostar masses to which this model applies is $M_{\star min} \approx 0.05\ M_\odot$ to $M_{\star max} \approx 10\ M_\odot$. The low end is arbitrarily chosen because the IMF is more uncertain at very low masses. The physical properties of low-and high-mass star formation are different, but it is unclear how significant these differences are, and what mass divides "low" from "high." Recent studies of spectral line maps indicates that the gas contracting to form a massive protostar behaves similarly to the gas in lower-mass systems, but with greater densities and velocities (Carolan et al 2009, Keto & Zhang 2010). In this paper, the maximum mass is set for masses where the mean column density starts to increase with radius, as discussed below.

*3.2. Infall duration and mass accretion rate*

The infall duration $t_f$ is calculated as a function of protostar mass $M_\star$ from equations (6)-(8). This duration increases from 0.01 Myr for objects with the mass of brown dwarfs, to 0.02 Myr for protostars with 0.1 $M_\odot$, to 0.1 Myr for solar-mass stars, and to 0.3 Myr for massive protostars, as shown in Figure 2. These durations span a factor 30 in time for a factor 200 in mass. Thus the infall durations have a shorter range than would be expected from a constant mass accretion rate. The maximum duration, 0.3 Myr, is significantly less than the typical cluster-forming life of 1-3 Myr. Such a significant time difference appears needed to allow massive stars enough time to form and to begin the large-scale dispersal of the cluster gas, as seen in W5 (Koenig et al 2008) and other regions with bubbles driven by massive stars (Churchwell et al 2006). The durations differ by less than a factor 2 between the IMFs of Kroupa (2002) and Chabrier (2005).

The instantaneous mass accretion rate increases with final protostar mass, as shown in Figure 3. As expected from the discussion in Section 2.3, at low protostar masses ~ 0.1 $M_\odot$ the instantaneous mass acccretion rate is nearly constant. This near constancy of the mass accretion rate, and its value of a few $10^{-6}$ $M_\odot$ yr$^{-1}$, resemble the infall of a singular isothermal sphere (Shu 1977) at initial temperature near 15 K. For high protostar mass approaching 10 $M_\odot$, the mass accretion rate increases linearly with mass and exceeds $10^{-4}$ $M_\odot$ yr$^{-1}$. Such high values have been estimated for massive young stars, based on observations of their outflows (Zhang et al 2005) and infall line profiles (Keto and Zhang 2010). Since the mass accretion rate is not constant, one should use the mean mass accretion rate $M_*/t_f$ rather than the instantaneous mass accretion rate $dM_*/dt$ to simply relate infall duration and final protostar mass.

*3.3. Density and column density profiles*

The density decreases with increasing radius, based on equations (13)-(17), as shown in Figure 4. The profiles in Figure 4 have similar shape for the two IMFs considered. The density decreases from a few $10^6$ cm$^{-3}$ to ~$10^4$ cm$^{-3}$ as the radius increases from 0.003 pc to 0.2 pc. At the smallest radii the log-log slopes approach -2, as for an isothermal self-gravitating body. As radius increases, the profile slope becomes progressively shallower.

The density profiles in Figure 4 can be represented conveniently by a sum of power laws,

$$n = Ar^{-v} + Br^{-w} \qquad (20)$$

where $v=2$ as in the "thermal-nonthermal" (TNT) models of Myers & Fuller (1992, hereafter MF). A similar treatment was given in the "two-component turbulent core model" (McKee & Tan 2003). MF used $w=1$ in an equilibrium model based on observed velocity dispersions. A better fit to the present profiles is obtained with $w=2/3$, with $A=34$ pc$^2$ cm$^{-3}$ and $B=2700$ pc$^{2/3}$ cm$^{-3}$ as shown in Figure 4. This fit value of $A$ corresponds to a singular isothermal sphere of temperature

$$T = \frac{m}{4k}\left(\frac{\pi G M_{\star n}}{\bar{\varepsilon} t_f}\right)^{2/3} \qquad (21)$$

where $k$ is Boltzmann's constant. For the adopted values of $\varepsilon = 0.5$ and $\bar{t}_f = 0.05$ Myr, the temperature is $T = 16$ K, similar to starless core temperatures in cluster-forming regions (Jijina, Myers & Adams 1999; Li, Goldsmith & Menten 2003).

In contrast, the density profile for a pressure-truncated isothermal sphere (Bonnor 1956, Ebert 1955, hereafter BE) cannot fit the profiles in Figure 4. As the radius increases, the log-log slope of the BE density profile becomes progressively steeper, whereas a good fit to Figure 4 would require the slope to become progressively shallower.

The radius where the thermal and nonthermal density components are equal gives a convenient definition of the "core radius" $r_{core}$. This is a fiducial radius which is useful for comparisons, but it does not correspond to a physical boundary. For the above fit parameters, equation (20) gives $r_{core} = 0.038$ pc, and the mass enclosed by this radius is 0.94 $M_\odot$. The efficiency with which this core makes the protostar mass at the mode of the IMF is 0.2. Note that these "core" properties refer to the core component of the condensation which makes a single protostar, as distinguished from the typical observed core.

Like the density in Figure 4, the mean column density has slope which becomes shallower with increasing radius, based on equations (13)-(15), (18) and (19), and as shown in Figure 5. The profiles for the two different versions of the IMF are once again similar, with mean column density decreasing from a few $10^{23}$ cm$^{-2}$ to ~$10^{22}$ cm$^{-2}$ as radius increases from ~0.003 pc to 0.2 pc. The mean column density within the core radius defined above is $2 \times 10^{22}$ cm$^{-2}$.

At radius exceeding 0.2 pc, and at corresponding protostar mass greater than 10 $M_\odot$, the mean column density begins to increase with increasing radius, which is inconsistent with observations of star-forming clouds. This discrepancy may indicate a mass above which the model is too idealized to be useful. The cold, spherical infall modelled in Section 2 does not account for the effects of hot, ionized gas (Keto 2003), and the isolated infall of a single condensation does not account for the high multiplicity of companions associated with OB stars (Zinnecker & Yorke 2007). Therefore the radius 0.2 pc and protostar mass 10 $M_\odot$ are taken as the maximum allowed values of the model.

*3.4. Accretion luminosity*

The distribution of spherical accretion luminosities is obtained from equations (2), (9), and (11). The peak of the distribution occurs at 4.5-5.6 $L_\odot$ depending on the IMF used, assuming $R_\star = 3\ R_\odot$. In three nearby star forming regions surveyed by the "c2d" program, 112 embedded protostars have a broad distribution of bolometric luminosities, with median 1.6 $L_\odot$ and quartile values 0.46 and 4.0 $L_\odot$ (Evans et al 2009, Enoch et al 2009, Dunham et al 2010). Matching the modal spherical accretion luminosity to the median bolometric luminosity requires the accretion luminosity efficiency $\gamma$ in equation (10) to have the value 0.29-0.36. This value suggests that the discrepancy between observed and model luminosities may be smaller than previously thought, as discussed in Section 4.2.

**4. Discussion**

This paper extends the basic idea of Paper 1, where differing protostar masses arise from condensations of similar structure, due to infalls of differing duration. This picture contrasts with the idea of constant star formation efficiency, where the mass of a protostar is proportional to the mass of the core where it forms. For cores whose boundaries are defined by a uniform background density, constant star formation efficiency requires their infalls to have essentially the same duration. In Paper 1 and in this paper, it is argued that such a narrow distribution of infall durations is not physically justified, and that a broader distribution such as the exponential distribution is more plausible.

This section discusses infall duration, initial density structure, and accretion luminosity, and compares to other models. Discussions about isolated and clustered star formation, low-mass and massive star formation, and the relation of the IMF to the observed core mass function are given in Paper 1.

*4.1. Infall duration*

The main assumptions of this paper are that protostar masses follow the IMF, that they arise from gravitational infall, and that their infall is equally likely to stop at any moment. The inferences about protostar formation times, mass accretion rate, and initial density structure are therefore independent of any *a priori* assumption about the dynamical status of the gas, about its equation of state, or about the nature of its turbulence.

On the other hand, the inferred properties depend on the assumed distribution of infall durations. Although no direct evidence about infall durations is available at present, it seems likely that their probability density distribution decreases with increasing duration, and has a breadth resembling the adopted waiting-time distribution.

The duration of infall probably varies over the ensemble of protostars in a crowded young cluster, a likely birthplace for stars which contribute to the IMF. There, star-forming gas can be dispersed by many causes, as discussed in Section 2.2.2 and in Paper 1. These opportunities for dispersal can be expected to occur over a range of times longer than the duration of a single outflow.

A distribution of infall durations much narrower than is assumed here cannot match observations, because then the accretion efficiency would become unphysically large. As noted in Section 2.5, the mass accretion efficiency $\varepsilon$ scales with the mean column density $\bar{N}$ and with the mean infall duration $\bar{t}_f$ as $\varepsilon \sim \bar{N}^{-3} \bar{t}_f^{-4}$. Thus if $\bar{N}$ were constant while $\bar{t}_f$ decreased by a factor 2, the accretion efficiency would increase by a factor $2^4=16$, to $\varepsilon = 8$. But this accretion efficiency cannot substantially exceed unity, since the model of cold, steady, spherical gravitational accretion gives the greatest possible mass accretion rate, in the absence of external compression.

The mean infall duration adopted here, $\bar{t}_f = 0.05$ Myr, is comparable to the duration of the "Class 0" stage of star formation, 0.1 Myr according to the c2d study of YSOs in nearby star-forming regions, after correction for extinction (Evans et al 2009, Enoch et al 2009, André, Ward-Thompson & Barsony 1993 (AWB)). It has been suggested that most of the protostar mass is accreted during this Class 0 phase, as opposed to the later Class I phase (AWB, White et al 2007). This idea is supported by a recent study of submillimeter continuum emission from 20 YSOs in nearby regions of low-mass star formation (Jørgensen et al 2009). In this group the densest "envelope" gas has median mass 1 $M_\odot$ for Class 0 sources, but only 0.1 $M_\odot$ for Class I sources. These results suggest that significant mass accretion is still underway for the Class 0 sources but has been largely completed for the Class I sources.

On the other hand, the Class 0 and Class I YSOs in the c2d sample have about the same mean luminosity, suggesting that the typical mass accretion rate is similar for each class. If so, the main period of mass accretion could be as great as the Class I lifetime, estimated as 0.5 Myr (Evans et al 2009, Enoch et al 2009). However, this conclusion is complicated by the broad spread in YSO bolometric luminosity, spanning three orders of magnitude within each class. This broad spread may be due to episodic accretion (Dunham et al 2010).

The question of lifetime estimation needs further investigation. For the present, it seems more direct to rely on the foregoing evidence from masses than from luminosities.

Further studies of infall histories are needed to refine our understanding of how infall stops. It would be useful to pursue simulations of stellar feedback in a cluster environment, as has been begun by Cunningham et al (2006), Carroll et al (2009), and Wang et al (2010), to study their effects on embedded star-forming condensations, and the resulting distributions of infall durations and protostar masses.

*4.2. Accretion luminosity*

As described in Section 3.4, the peak of the distribution of model accretion luminosities matches the peak of the observed distribution of bolometric luminosities, 1.6 $L_\odot$, provided the accretion luminosity efficiency $\gamma$ is 0.29-0.36, as described in Section 3.4. The corresponding factor *$1/\gamma$* =2.8-3.4 is substantially less than the factor 10-50 discrepancy between predicted and observed luminosities of protostars in Taurus (Kenyon & Hartmann 1995, hereafter KH). These predicted luminosities were based on a model of constant accretion rate (Shu 1977), assuming a protostar mass 0.4 $M_\odot$, typical of stars on the birthline in Taurus. A similar discrepancy is noted with the more recent data of the c2d program (Evans et al 2009, Enoch et al 2009, Dunham et al 2010).

With the present model, the discrepancy is reduced because the mass which provides the peak of the luminosity distribution is 0.11-0.12 $M_\odot$. These low values arise because the luminosity distribution is obtained from the IMF, whose modal mass is 0.16-0.20 $M_\odot$ as given in the Appendix. Further, the mass for the mode of the luminosity distribution is less than the mass for the mode of the IMF, because in this model the mass accretion rate increases with mass. Thus the peak of the luminosity distribution arises from a mass nearly four times less than the typical mass assumed by KH.

The remaining factor of ~3 discrepancy may be explained by episodic accretion (KH, Vorobyov & Basu 2005, Dunham et al 2010), a proposal which can be confirmed by observations of sufficiently luminous bursts. In addition, the discrepancy may arise simply from the nonspherical nature of the accretion, since the values of $\gamma = 0.3–0.4$ derived here are only slightly less than the disk accretion values $\gamma = 0.4$-$0.5$ considered by BCG.

The present comparison uses the mode of the observed and predicted luminosity distributions, rather than the mean, since the mode is independent of the range of luminosities considered. However a stronger constraint can be obtained by comparing the observed and predicted distributions themselves. Such comparison would be useful for the future, but is beyond the scope of this paper.

*4.3. Core-clump density structure*

The density structure derived from an exponential distribution of infall durations and the IMF in Section 3.3 resembles several descriptions of observed "cores" having steep density gradients in "clumps" having shallower density gradients, in nearby star-forming regions (Kirk, Johnstone, & Di Francesco 2006, Bergin & Tafalla 2007). This agreement with observations suggests that the exponential distribution is a useful description of infall durations.

The exponential distribution is not the only distribution of infall durations tied to core-clump density structure. The assumption of "equally likely stopping" in Section 2.2.2 requires that the probability density of infall stopping $s(t_f)$ be constant. A similar but less restrictive assumption is "no preferred stopping time" so that $s(t_f)$ varies monotonically with infall duration but has no local maximum. Then calculations similar to those in Section 3 indicate that the exponential distribution is one of a family of distributions which match the IMF and which yield a core-clump density profile.

Both the core and the clump components of the core-clump profile appear necessary. If a single power-law density profile $n \sim r^{-\alpha}$ is assumed, matching the IMF of Kroupa (2002) requires $s(t_f)$ to have a local maximum, i.e. a preferred stopping time, unless the exponent $\alpha$ exceeds $\alpha_{min} \equiv 6(1-b)/(3-2b)=1.75$. Here $b=0.3$ as given in the Appendix.

This result excludes a shallow-slope clump which lacks an embedded core. Such a clump with $\alpha=2/3$ as in Section 3.3 would require the distribution of infall durations to have a preferred stopping time. Similarly, a core with $\alpha=2$ having no associated clump is also excluded, because for massive stars its mean column density is too low and its accumulation length is too great. For $\varepsilon = 0.5$ and $\bar{t}_f = 0.05$ Myr, as assumed earlier, the initial condensation for a 10 $M_\odot$ protostar must have mean column density $\bar{N} = 6 \times 10^{20}$ cm$^{-2}$ and diameter $d = 1.5$ pc, whereas most nearby clusters have $\bar{N} > 10^{22}$ cm$^{-2}$ and $d < 1$ pc (Gutermuth et al 2009).

Thus the core-clump density structure appears to follow generally from the requirements to match the IMF with no preferred infall duration, and to make massive stars in clumps whose column density and extent match those of observed cluster-forming regions.

*4.4. Relation to equilibrium core models*

The steep-slope portion of the core-clump profile resembles the structure of the thermally supported SIS. However the SIS is a limiting case of the physically more realistic nonsingular isothermal sphere (Chandrasekhar 1939). The nonsingular isothermal sphere is nearly uniform inside its first thermal scale length, and many isolated cores show such "flat-top" density structure (Ward-Thompson et al 1994, Di Francesco et al 2007). A calculation similar to those above for the IMF of Kroupa (2002) and for the density profile of a nonsingular isothermal sphere requires a sharp local maximum in its distribution of infall durations, due to this nearly uniform density zone. As discussed earlier, such a sharp maximum is unlikely if the infall is terminated by processes of various start times and durations.

Thus if the typical initial condensation is close to isothermal equilibrium on small scales, this equilibrium must be indistinguishable from its SIS limit for the range of masses considered. Such equilibrium is possible if the mass within the first thermal scale length is less than the adopted mass lower limit, $\varepsilon^{-1}M_{\star min} = 0.1\ M_\odot$. Such a small scale length can arise if the central density is sufficiently high and/or if the central temperature is sufficiently low.

Both of these conditions may be met in core interiors in cluster-forming regions. For the temperature 16 K estimated in Section 3.3, the corresponding minimum value of the peak density is $5 \times 10^5$ cm$^{-3}$. This minimum density is probably exceeded by many cores in the Perseus complex, where the mean density exceeds $3 \times 10^5$ cm$^{-3}$ for 103 cores (Hatchell & Fuller 2008). Further, the gas temperature may decrease toward the position of greatest column density due to absorption of externally incident heating photons (Evans et al 2001). The densest part of a starless core in the cluster-forming region Ophiuchus D has central temperature ~ 6 K according to observations of the $1_{10}$-$1_{11}$ line of H$_2$D$^+$. This temperature is significantly lower than estimates based on lower-density tracers (Harju et al 2008).

*4.5. Initial conditions*

Despite the uncertainty in the distribution of infall durations, the predicted properties agree well with observational constraints. The predicted infall times for protostars of mass 0.05 $M_\odot$ to 10 $M_\odot$ are 0.01 to 0.3 Myr. These times lie well within the few Myr star-forming life of a cluster (Muench et al 2007, Evans et al 2009). These relative ages allow multiple generations of low-mass stars, and also allow for significant feedback from more massive stars leading to the large-scale dispersal of the cluster gas. The mass accretion rates match observational estimates for protostars of low mass within a factor of a few (Evans et al 2009) and of high mass within an order of magnitude (Zhang et al 2005). The typical condensation has a density profile matching the core-clump structure as observed in nearby star-forming regions (Kirk, Johnstone & Di Francesco 2006, Bergin & Tafalla 2007).

The core and clump components are warmer and denser than their counterparts in regions of isolated star formation. The effective core temperature is ~ 16 K, greater than the temperature ~10 K seen in more isolated regions (Jijina, Myers, & Adams 1999). The clump density exceeds $10^4$ cm$^{-3}$, typical of cluster-forming gas, and greater than clump densities of ~$10^3$ cm$^{-3}$ in more isolated regions (Bergin & Tafalla 2007).

*4.6. Spherical and filamentary geometry*

The core-clump structure inferred for spherical geometry could take more than one 3D form. Paper 1 showed that centrally condensed environments, self-gravitating isothermal filaments or isothermal layers, can give mass functions which resemble the IMF, provided their peak density is great enough.

Filaments which are less centrally condensed than the isothermal filament may provide an environment which can feed mass to a protostar at a greater rate than can the isothermal filament, which is strongly centrally condensed. An example of such structure is the filamentary dark cloud L977. There, extinction star counts indicate that the density profile in the direction *r* perpendicular to the symmetry axis decreases as $r^{-p}$, where *p* is much closer to 2 than to the value 4 expected for the outer parts of a self-gravitating isothermal filament (Alves et al 1998). Similarly, structures which are intermediate between a filament and a layer can provide more infalling mass in a given time than can a self-gravitating isothermal filament alone (Schmid-Burgk 1967, Myers 2009a).

*4.7. Comparison to other models*

The exponential waiting-time distribution of infall durations in this paper resembles those used by Myers (2000), Basu & Jones (2004), Bate & Bonnell (2005), and Paper 1. As discussed in Paper 1, this concept has a physical justification which is more plausible than assuming that all infalls have essentially the same duration.

The two-component initial density structure obtained here is similar to the "TNT" models of Myers & Fuller (1992) and Caselli & Myers (1995), and to the "two-component turbulent core" models of McKee & Tan (2003). In turn the TNT models are an extension of the SIS model (Shu 1977).

Each of these two-component density models was posited to account for massive stars in a picture similar to that for low-mass stars. However in this paper the two-component character is not assumed, and is not obtained from a model of equilibrium structure. Instead it is a consequence of assuming that protostar final masses follow the IMF, and that their infall has equal likelihood of stopping at any moment.

The reliance of the present model on a distribution of infall times resembles the IMF model of Adams & Fatuzzo (1996), where distributions of several input variables are considered. The two models are also similar because each relies on outflows to help terminate the star-forming infall. The models differ in their predictions of infall durations, since Adams & Fatuzzo (1996) predict durations more narrowly concentrated around 0.1 Myr than does the present model.

Some of the foregoing models are expressed in a common framework of mass accretion rate, and are compared with the competitive accretion model of Bonnell et al (1997), in a recent paper by McKee & Offner (2009, hereafter MO). For comparison between models, temperature 16 K and mean density $9 \times 10^4$ cm$^{-3}$ are assumed, based on the core-clump model in this paper for a 1 $M_\odot$ protostar.

With these assumed values, the time to form a 1 $M_\odot$ protostar is 0.1 Myr for the core-clump model, for the two-component turbulent core, and for the competitive accretion models, based on the model descriptions in MO. The corresponding time is 0.3 Myr for the SIS accretion model. These formation times are similar, probably because the models have similar mass accretion rates for relatively low masses. It will be useful to compare these models in more detail, especially for formation of more massive protostars.

## 5. Summary

The main features of this paper are:

1. A spherically symmetric condensation forms a single protostar, whose final mass scales with the condensation mass available in a free-fall time. The scaling factor is the mass accretion efficiency $\varepsilon$, an adjustable parameter.

2. The final protostar mass follows the probability distribution set by the IMF. Continuous approximations are made to the segmented IMF representations of Kroupa (2002) and Chabrier (2005).

3. The infall duration follows the exponential distribution, a waiting-time probability distribution with equally likely stopping. This property describes conditions in a young cluster, where dense gas is dispersed asynchronously by stellar feedback, turbulent flows, competitive accretion, and dynamical ejection. The mean infall duration $\bar{t}_f$ is an adjustable parameter.

4. The adopted parameter values $\varepsilon = 0.5$ and $\bar{t}_f = 0.05$ Myr match the typical column density $10^{22}$ cm$^{-2}$ averaged over radius 0.2 pc in nearby young clusters. The model applies to protostar masses in the range 0.05 $M_\odot$ to 10 $M_\odot$. All calculated properties differ by a factor less than 2 between the IMF representations of Kroupa (2002) and Chabrier (2005).

5. Protostar formation times range from 0.01 to 0.3 Myr for protostar masses 0.05 to 10 $M_\odot$. These times are short enough to allow multiple generations of protostars in the typical star-forming life of a young cluster.

6. The protostar mass accretion rate increases from a few $10^{-6}$ $M_\odot$ yr$^{-1}$ at low mass, to a few $10^{-4}$ $M_\odot$ yr$^{-1}$ at high mass, matching observational estimates for YSOs of low and high mass (Evans et al 2009, Keto & Zhang 2010). At low mass the mass accretion rate is nearly constant, as expected for collapse of a primarily thermal condensation. At high mass the mass accretion rate increases linearly with mass, as expected from the power-law nature of the high-mass IMF, and from the assumption of equally likely stopping.

7. The typical model luminosity due to spherical accretion exceeds the typical bolometric luminosity of protostars in nearby regions by a factor ~3. This discrepancy is less than the factor 10-50 "luminosity problem" first noted by Kenyon & Hartmann (1995). The discrepancy is reduced because the protostar mass which gives the peak of the luminosity distribution is nearly 0.1 $M_\odot$, less than the protostar mass at the peak of the IMF, and less than the protostar mass used for earlier comparisons. The remaining discrepancy may be accounted for by the nonspherical and episodic nature of the accretion.

8. The typical initial density profile declines steeply on small scales and shallowly on large scales, resembling the character of "cores" in "clumps" observed in nearby star-forming regions. The density profile is fit by a sum of power laws of radius, as in equation (20) and Figure 4. The core component is a SIS with temperature 16 K, and the clump component decreases with increasing radius $r$ as $r^{-2/3}$. The "core radius" where the two density components are equal is 0.04 pc, which encloses mass 0.9 $M_\odot$ and mean column density $2 \times 10^{22}$ cm$^{-2}$.

9. The steep central density profile is consistent with the uniform zone of a nonsingular isothermal equilibrium sphere, only if the central density and temperature provide a sufficiently small thermal scale length. These conditions may be met if the central density typically exceeds $\sim 5 \times 10^5$ cm$^{-3}$ for central temperature $< \sim 16$ K.

10. The core-clump density structure resembles the "TNT" model of Myers & Fuller (1992) and the "two-component turbulent core" model of McKee & Tan (2003). Both the core and the clump components appear necessary to match both the IMF and the gas properties of young clusters.

**Appendix. Continuous approximations to the initial mass function**

An approximation to the IMF of Kroupa (2002) has probability density

$$p_K(M_\star) = \frac{a}{M_{\star n} \mu^b (1 + \mu^c)} \qquad (A1)$$

where the normalizing mass scale is $M_{\star n} = 0.205\ M_\odot$, and where $\mu$ is the dimensionless protostar mass,

$$\mu \equiv \frac{M_\star}{M_{\star n}} \quad . \tag{A2}$$

Here $b = 0.30$ and $c = 2.05$. The constant $a = 0.636$ is obtained from equation (A1) by requiring that the integral of the probability density over all masses be unity.

Similarly, a continuously differentiable approximation to the probability density for the IMF of Chabrier (2005) is

$$p_C(M_\star) = d \frac{\text{erfc}(g - h \ln \mu)}{M_{\star n} \mu^f} \tag{A3}$$

following the expression given in Basu & Jones (2004). Here $d = 0.846$, $f = 2.45$, $g = 0.620$, and $h = 0.655$.

**Acknowledgements** Helpful discussions and comments were provided by Paola Caselli, Mark Heyer, Helen Kirk, Charlie Lada, Chris McKee, Tom Megeath, Stella Offner, Frank Shu, and Qizhou Zhang. The referee, Fred Adams, provided numerous suggestions which improved the content and the clarity of the presentation. Support and encouragement from Irwin Shapiro and Terry Marshall are gratefully acknowledged.

McKee, C., & Tan, J. 2003, ApJ, 585, 850

McKee, C., & Offner S. 2009, ApJ, submitted

Motte, F., André, P., & Neri, R. 1998, A&A, 336, 150

Muench, A., Lada, C., Luhman, K., Muzerolle, J., & Young, E. 2007, AJ 134, 411

Myers, P., & Fuller, G. 1992, ApJ 396, 631

Myers, P. 2009a, ApJ, 700, 1609

Myers, P. 2009b, ApJ, 706, 1341 (Paper 1)

Nadarajah, S. 2007, Comput. Ind. Eng, 53, 693

Nutter, D., Ward-Thompson, D., & André, P. 2006, MNRAS, 368, 1833

Reipurth, B., & Clarke, C. 2001, AJ, 122, 432

Salpeter, E. 1955, ApJ, 121, 161

Sandell, G., & Knee, L. 2001, ApJ 546, 49

Schmid-Burgk, J. 1967, ApJ 149, 727

Shu, F. 1977, ApJ, 214, 488

Shu, F. 1992, *The Physics of Astrophysics, Vol. II: Gas Dynamics* (Mill Valley, CA: Univ. Science Books)

Stahler, S. 1988, ApJ 332, 804

Stanke, T., & Williams, J. 2007, AJ, 133, 1307

Stüwe, J. 1990, A&A 237, 178

Swift, J., & Williams, J. 2008, ApJ, 679, 552
32

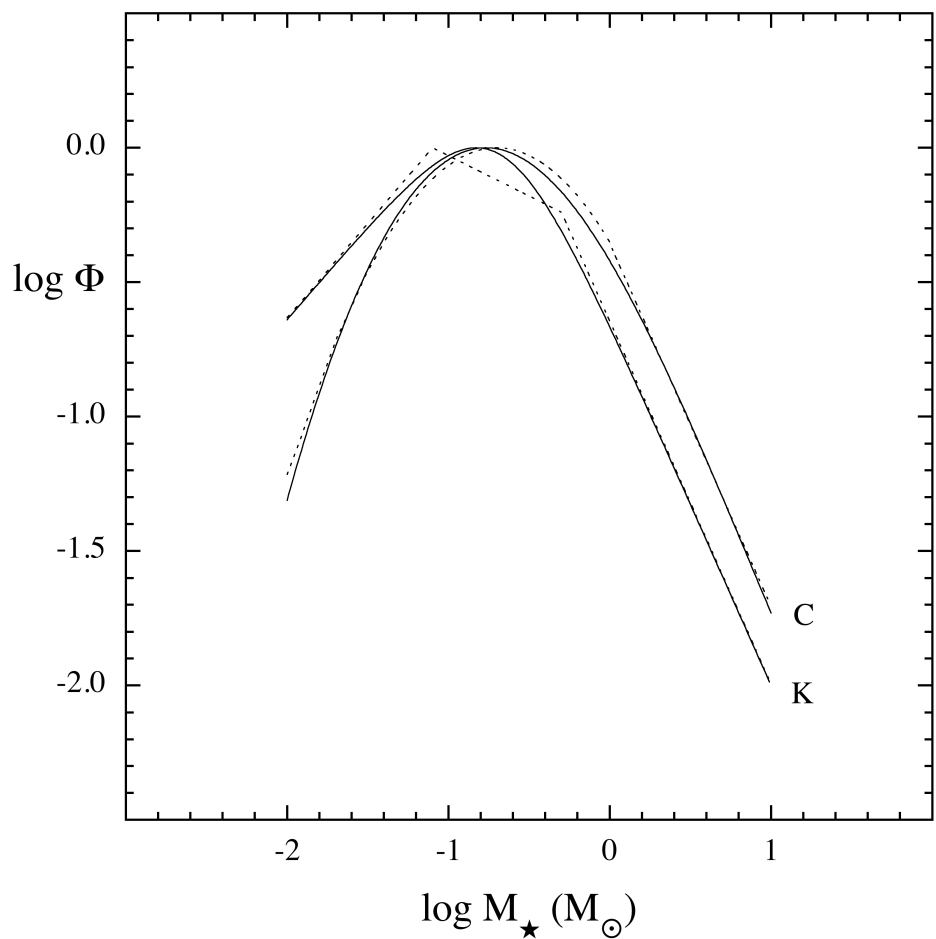

**Figure 1.** Initial mass functions of Kroupa (2002, *K*) and Chabrier (2005, *C*), shown as dotted lines, and their continuous approximations, given in the Appendix, shown as solid lines. Each function has been normalized to a maximum value of unity.

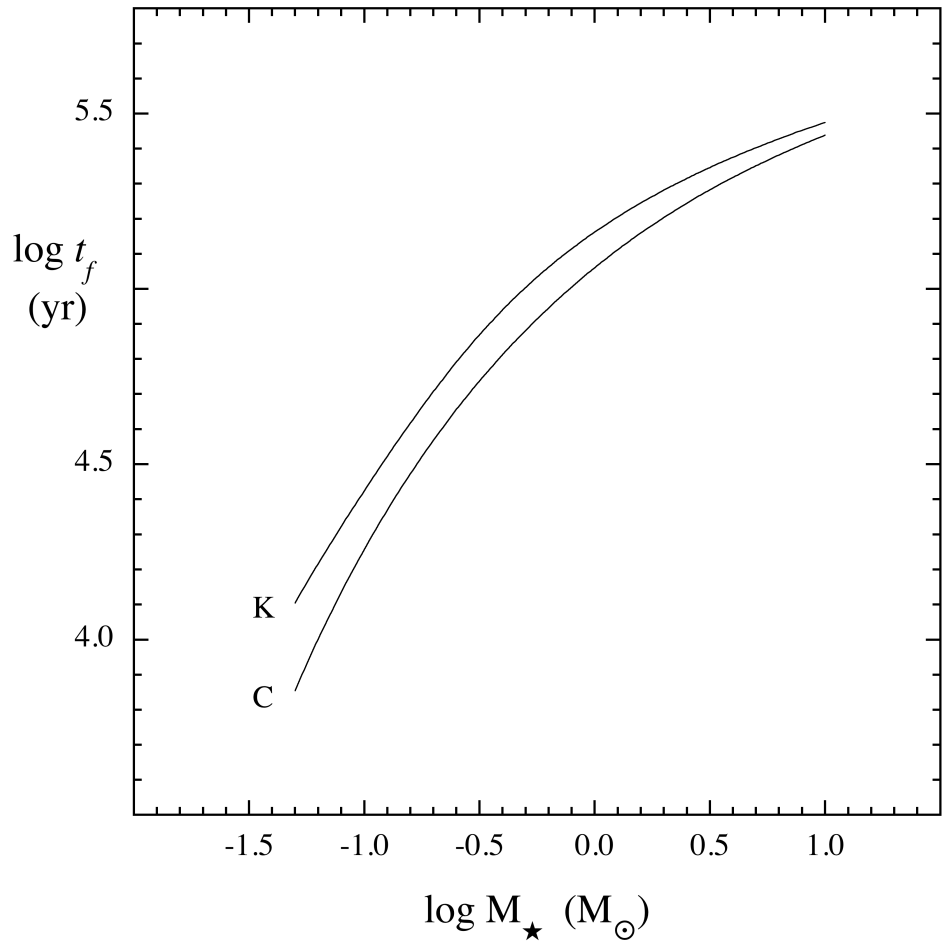

**Figure 2.** Infall duration $t_f$ for protostars of final mass $M_\star$ based on mean infall time $\bar{t}_f = 0.05$ Myr and on the IMFs of Kroupa (2002, *K*) and Chabrier (2005, *C*).

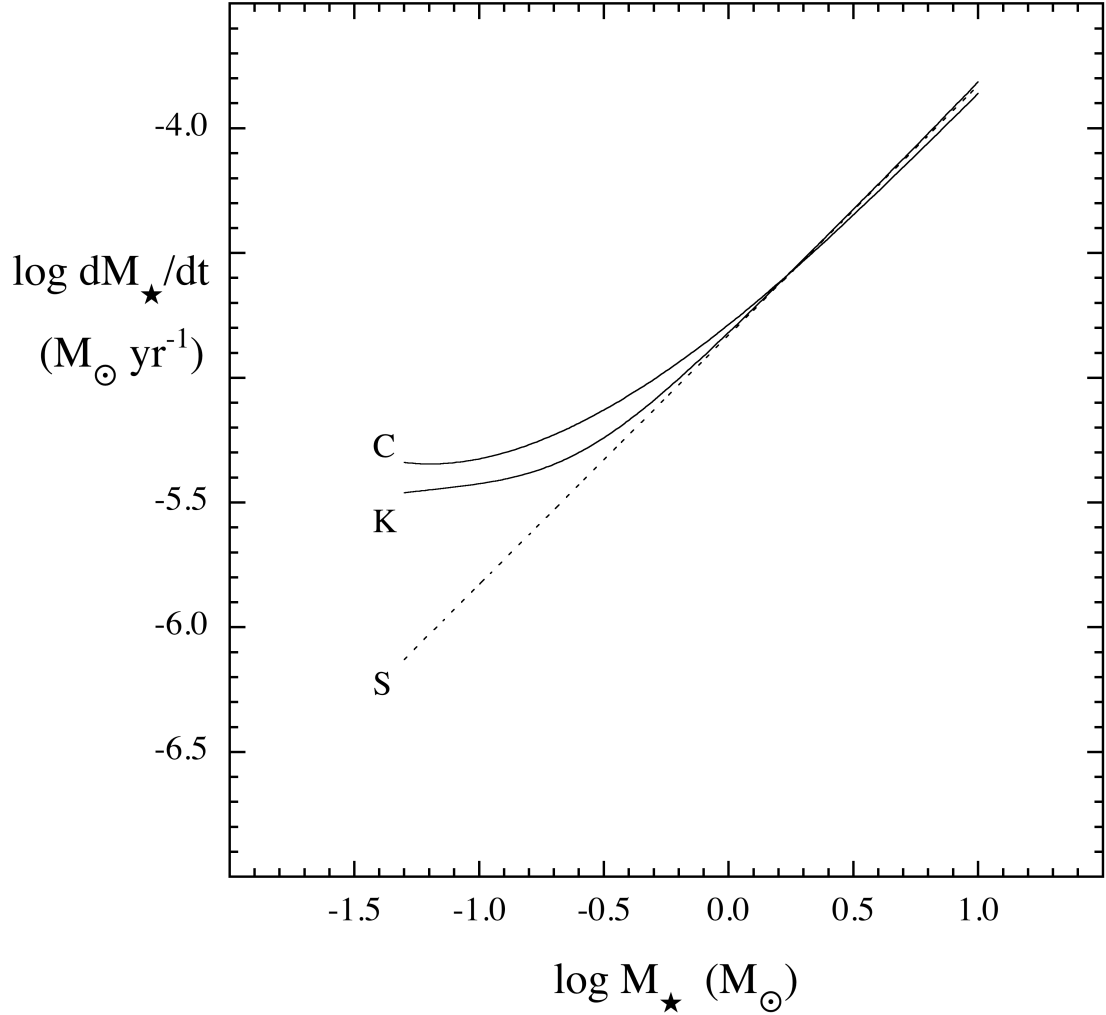

**Figure 3.** Mass accretion rate $dM_\star/dt$ when a protostar has reached its final mass $M_\star$, based on the IMFs of Kroupa (2002, $K$), Chabrier (2005, $C$), and Salpeter (1955, $S$). The mean infall stopping time is $\bar{t}_f = 0.05$ Myr. For IMFs having a low-mass turnover and a high-mass power law ($K$, $C$), the rates are nearly constant at low mass and increase as $M_\star$ at high mass. For an IMF of the power-law form $\Phi \sim M_\star^{-s}$, the dotted line ($S$) shows the relation $dM_\star/dt = M_\star/(s\,\bar{t}_f)$, where the log-log slope has the Salpeter value $s=1.35$.

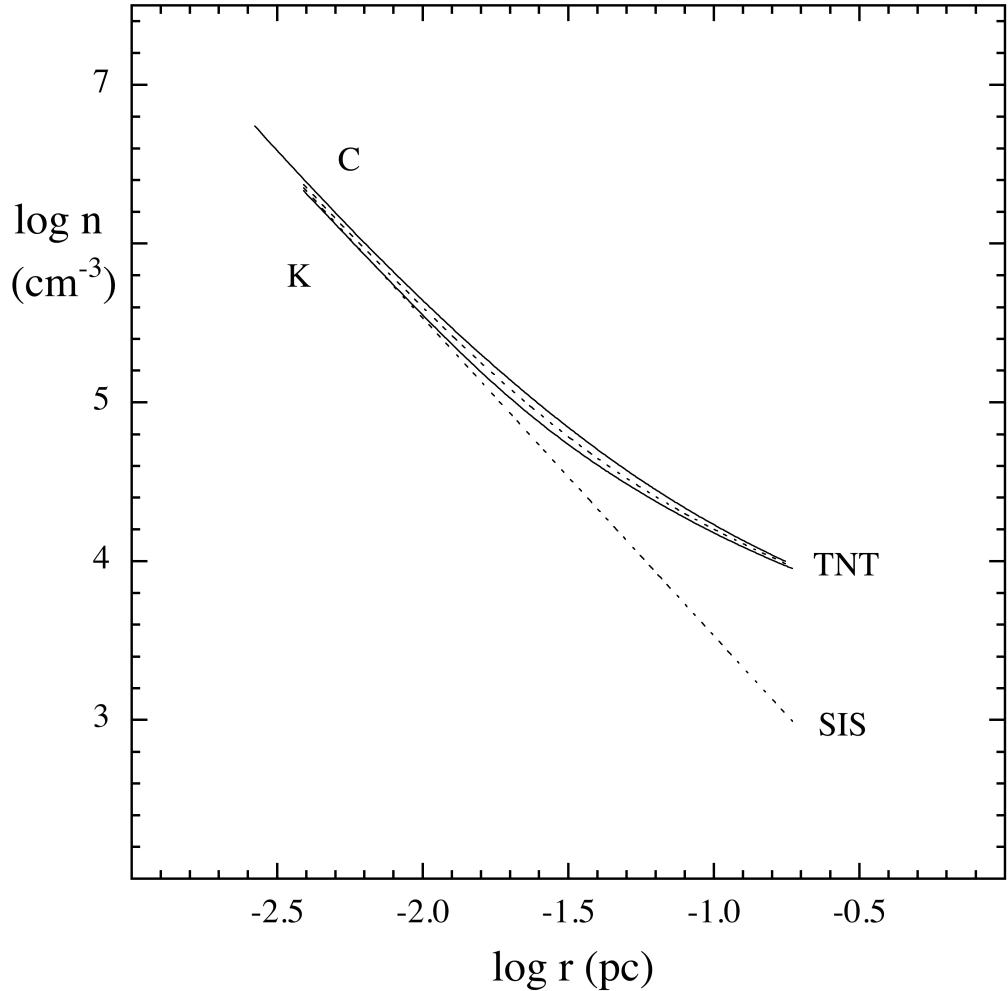

**Figure 4.** Initial gas density $n$ as a function of spherical radius $r$, based on the IMFs of Kroupa (2002, $K$) and Chabrier (2005, $C$), and on accretion efficiency $\varepsilon = 0.5$, and mean infall time $\bar{t}_f = 0.05$ Myr. At small radii, these density profiles are well fit by a singular isothermal sphere at temperature 16 K *(dotted line, SIS)*. At larger radii, two components are needed *(dotted line, TNT)*.

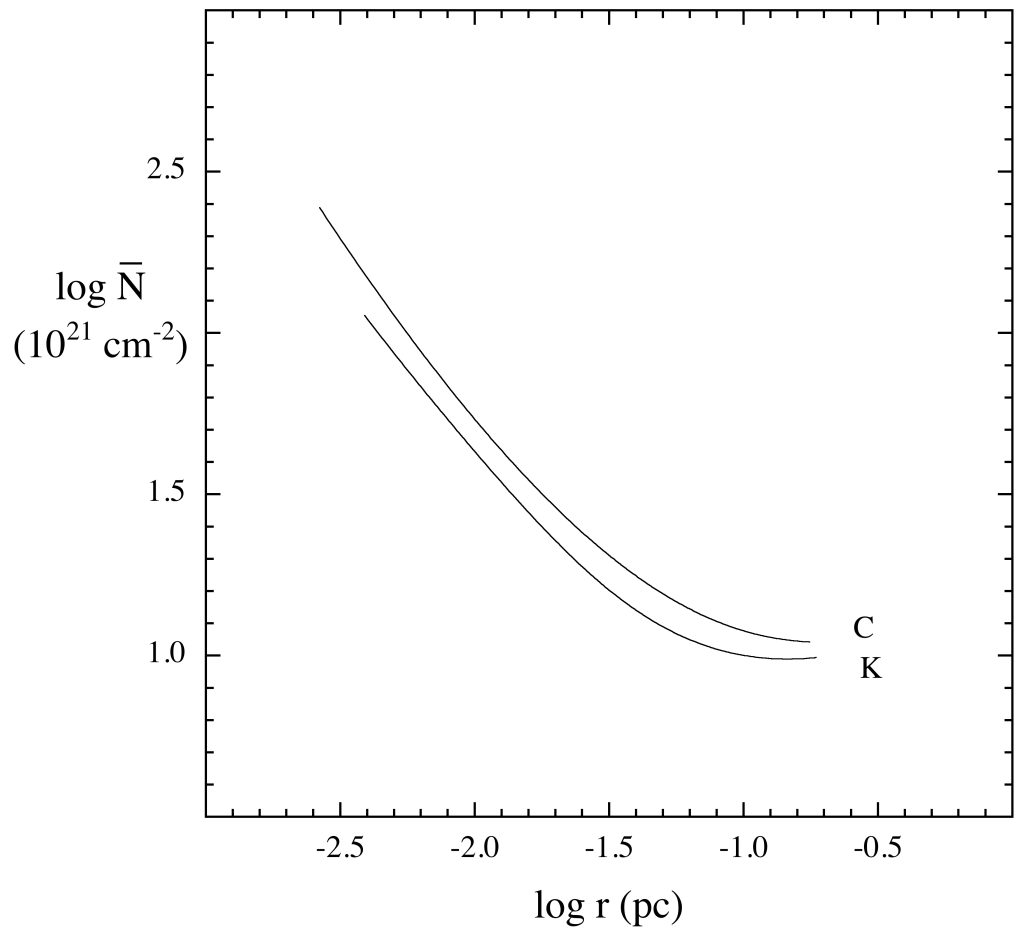

**Figure 5.** Initial mean column density within spherical radius $r$, as a function of $r$, based on the IMFs of Kroupa (2002, $K$) and Chabrier (2005, $C$), and on accretion efficiency $\varepsilon = 0.5$, and mean infall time $\bar{t}_f = 0.05$ Myr.